\documentclass{llncs}

\usepackage{amsmath}
\usepackage[ansinew]{inputenc}
\usepackage[matrix,arrow,curve,frame]{xy}
\usepackage{url}
\usepackage[dvips]{graphicx}
\usepackage{subfigure} 
\usepackage{multicol}
\usepackage{pstricks}
\usepackage{epsfig}

\newcommand{\mr}[1]{{\mathrm{#1}}}
\newcommand{\ket}[1]{\left | \, #1 \right \rangle}

\newcommand{\braket}[2]{\left \langle #1 \, | \, #2 \right \rangle }
\newcommand{\ketbra}[2]{\left | \, #1 \rangle \langle #2 \, \right | }

\newcommand{\modulop}[2]{\left | #1 \right |^{#2}}

\newcommand{\urlbib}[1]{{\footnotesize{\url{#1}}}}

\begin{document}
\title{Computing a Turing-Incomputable Problem from Quantum Computing} 

\author{Andrés Sicard \and Mario Vélez \and Juan Ospina}
\institute{
Departamento de Ciencias Básicas, Universidad EAFIT, A.A. 3300
\\
Medellín, Colombia
\\
\email{\{asicard, mvelez\}@eafit.edu.co} \email{judoan@epm.net.co}
}

\maketitle

\begin{abstract}
A hypercomputation model named \emph{Infinite Square Well Hypercomputation Model  (ISWHM)} is built from quantum computation. This model is inspired by the model proposed by Tien D. Kieu \cite{Kieu-2003a} and solves an Turing-incomputable problem. For the proposed model and problem, a simulation of its behavior is made. Furthermore, it is demonstrated that ISWHM is a universal quantum computation model.
\end{abstract}


\section{Introduction}
In general terms, a quantum computation model is built based on a physical re\-fe\-rent which determines the mathematical and computational characteristics of the quantum algorithms built on such model. These algorithms, in turn, determine the kinds of problems that can be solved under the quantum computation model. 

The first quantum computation model considered as a hypercomputation model was the one proposed by Tien D. Kieu \cite{Kieu-2003a} having the following characteristics: (i) Physical referent: quantum harmonic oscillator, (ii) Turing-incomputable problem that solves: Hilbert's tenth problem.

This article presents the construction of a hypercomputation model named \emph{Infinite Square Well Hypercomputation Model (ISWHM)}. This model is based on the one proposed by Kieu, but we have selected the infinite square well instead of the quantum harmonic oscillator as the physical referent. This change conveys a change in the Turing-incomputable problem to solve, and instead of solving Hilbert's tenth problem it solves a problem equivalent to it. A simulation of its behavior is carried out for the proposed model and problem. Besides, it is demonstrated that ISWHM is a universal quantum computation model.

\section{Hypercomputation}\label{secc-hipercomputacion}
While the idea of an \emph{absolute computability}, detached from logical, mathematical, physical or biological theories is hard to support nowadays, the idea of a \emph{relative computability} has progressively gained supporters, as shown by the establishment of an academic community around the topic (\url{www.hypercomputation.net}). 

The term `hypercomputer' or `hypermachine' denotes any data processing device (theoretical, potencially realizable or that can be implemented) capable of carrying out tasks that cannot be performed by a Turing machine \cite{Copeland-Proudfoot-1999a}. A hypermachine, also capable of simulating a universal Turing machine is called a super-(Turing machine), otherwise it is called a non-(Turing machine) \cite{Stannett-2003}.

At first glance, we could think that the possibility of a hypercomputation model would be a refutation of the widely accepted Church-Turing thesis, which identifies the \emph{naturally calculable} functions with the Turing-computable functions. But actually the existence of hypercomputation models refute the $M$ thesis, which identifies the functions \emph{calculable by a machine} with the Turing-computable functions \cite{Sicard-Velez-2001a}. On the other hand, notwithstanding the proliferation of theoretical hypercomputation models \cite{Ord-2002}, the possibility of the \emph{real} construction of a hypermachine keeps being controversial and under analysis.

\section{The Turing-Incomputable Problem}\label{secc-problemaNoTuring}
The demonstration that ISWHM is a hypercomputation model will be done by demonstrating that it can solve an Turing-incomputable problem. 

Let $\bbbn$ be the set of non-negative integers, $\bbbz$  the set of the whole numbers,  $\bbbz^+$ the set of the positive integers, and  $\bbbz^{+2}$ the set of the perfect square numbers. A Diophantine equation is one of the form

\begin{equation}\label{asicard-eq-10}
D(x_1, \dots, x_k) = 0 \enspace ,   
\end{equation}
where $D$ is a polynomial with integer coefficients. In present terminology, Hilbert's tenth problem may be paraphrased as: Given a Diophantine equation of type \eqref{asicard-eq-10}, we should build a procedure to determine whether or not this equation has a solution in  $\bbbz$. From the concluding results gotten by Matiyasevich, we know that, in the general case, this problem is algorithmically insolvable or more precisely, it is Turing-incomputable.

Having into account that any non-negative integer number may be expressed as the addition of the squares of four integer numbers (the four squares theorem, due to Lagrange), and that any system of a finite number of Diophantine equations has a solution in  $\bbbz$, if and only if a certain Diophantine equation associated to the system has a solution in  $\bbbz$; the Hilbert's tenth problem may be reduced to search a solution in $\bbbn$ \cite{Matiyasevich-1993}.

Due to the Hamiltonian operator \eqref{cp-refesimu-15} used in the description of ISWHM, the decision problem associated with this model of computation consists in determining whether or not a Diophantine equation kind \eqref{asicard-eq-10}  has a solution in $\bbbz^{+2}$. However, this problem is equivalent to determining whether or not the Diophantine equations system 

\begin{equation*}
  D(x_1, \dots, x_k) = 0, \quad x_1 = y_1^2, \quad \dots \quad x_k = y_k^2 \enspace ,
\end{equation*}
has a solution in $\bbbz$, and this problem is equivalent to Hilbert's tenth problem. Therefore, the decision problem associated with ISWHM is a Turing-incomputable one.

\section{The Physical Referent}
The energy levels  $E_{n}$, for a particle with mass $m$ in a infinite square well with length $L$, are given by \cite{Galindo-Pascual-1978}

\begin{equation*}
 E_n=\frac{\hbar^2\pi^{2}}{2mL^{2}}n^{2} \enspace .
\end{equation*}

These energy levels are the eigenvalues of the Hamiltonian operator of the system denoted by $H$, for which correspond as eigenvectors the states denoted by the kets $\ket{n} $, where $n \in \bbbz^{+}$. The Hilbert space for the infinite square well, denoted $\bbbh_{\mr{c}}$, is infinite dimensional, and its orthonormal and complete canonical base is the countably infinite set $\{\ket{n}\}_{n  \in \bbbz^{+}}$. In such base, $H$ is diagonal and has the following form 

\begin{equation} \label{cp-refesimu-15}
 H\ket{n}=\frac{\hbar^2\pi^{2}}{2mL^{2}}n^{2}\ket{n} \enspace .
\end{equation}

If matrix $M$ is defined as

\begin{equation*}
 M=\frac{2mL^{2}}{\hbar^2\pi^{2}}H \enspace ,
\end{equation*}
then \eqref{cp-refesimu-15} may be written 

\begin{equation}\label{cp-refesimu-25}
 M\ket{n}=n^{2}\ket{n} \enspace .
\end{equation}

Equation \eqref{cp-refesimu-25} implies that matrix $M$ has a diagonal form in its canonical base $\{\ket{n}\}_{n  \in \bbbz^{+}}$  and its components are the  $\bbbz^{+2}$ numbers. This canonical base satisfies the following completeness and orthonormality conditions

\begin{equation*}
 \sum_{n=0}^{\infty}\ketbra{n}{n} = \bbbone, \quad \braket{n}{m} = \delta_{n,m} \enspace ,
\end{equation*}
where $\bbbone$  is the identity matrix in Hilbert space $\bbbh_{\mr{c}}$.

With this mathematical description of the infinite square well we may proceed to the extension of Kieu's model from the quantum harmonic oscillator towards the infinite square well thus to obtain the ISWHM.

\section{The Algorithm}
If we have a Diophantine equation of type \eqref{asicard-eq-10}, the strategy to determine whether or not it has a solution in $\bbbz^{+2}$, inspired in the strategy followed by Kieu, consists in translating equation $D$ in $k$ variables into a codifying Hamiltonian operator  $H_D$ represented by a countably infinite dimensional matrix which acts in Hilbert space  $H_{\mr{c}}^{\otimes k}=\underbrace{H_{\mr{c}} \otimes \dots \otimes H_{\mr{c}}}_{k-\mr{times}}$, and then restate the problem in terms of the $H_D$ properties \cite{Kieu-2003a}. 

Particularly, each of the $k$ variables is substituted in $D$ by the operator $M$ given by \eqref{cp-refesimu-25} and we obtain the Hamiltonian $H_{D}$ given by 
\begin{equation*}
  \label{cp-refesimu-65}
 H_{D}=D(M_{1},M_{2}, \dots ,M_{k }) ^{2} \enspace ,
\end{equation*}
and therefore, equation $D$ has at least a solution in $\bbbz^{+2}$, if and only if the eigenvalue $E_0$ associated to the $H_{D}$ fundamental state is zero.

The problem of determining the $E_0$ eigenvalue may be solved by the application of the adiabatic theorem, as Kieu does in the harmonic oscillator case. For this doing a universal interaction Hamiltonian operator  $H_{\mr{I}}$ (valid for all Diophantine equations) is introduced, which describes an operator that acts in Hilbert space  $\bbbh_{\mr{c}}^{\otimes k}$  and has a eigenvector known in such a space that corresponds to the zero eigenvalue.

The Hamiltonian operator $H_{\mr{A}}(t/T)$ that results from the convex overlapping of $H_{D}$ and $H_{\mr{I}}$, with adiabatic parameter $t/T$ has the form 

\begin{equation*}
  \label{cp-refesimu-70}
 H_{\mr{A}}(t/T)=(1-t/T)H_{\mr{I}} +  (t/T)H_{D} \enspace ,
\end{equation*}
where $T$ is the total time of the adiabatic evolution and $t \in [0,T]$. The Hamiltonian
 $H_{\mr{A}}(t/T)$ determines the solution of the Schrödinger discretized equation given by \cite{Kieu-2003}

\begin{equation}
  \label{cp-refesimu-90}
 \psi(t+\delta t)=\left ( \frac{1-\frac{i}{2}H_{\mr{A}}(t/T)\delta t}{1+\frac{i}{2}H_{\mr{A}}(t/T)\delta t} \right ) \psi(t)\enspace ,
\end{equation}
where $\delta t$ is the time step.

For a Diophantine equation $D(x)=0$, based on \eqref{cp-refesimu-90} the algorithm for ISWHM to determine whether or not it has a solution in $\bbbz^{+2}$, has as inputs and outputs the ones indicated in Fig. \ref{fig05}, where for $\psi(t) = \sum_n c_n(t) \ket{n}$:

\begin{enumerate}
\item $P_n(t)=\modulop{c_n(t)}{2}$ represents the probabilities of the quantum states  $\ket{n}$ for a time $t$.

\item $<n^2(t)> =  \sum_{n} P_n(t) n^2$ represents the expected value of the square random variable $n$ that codifies the
    value of the unknown $x$ in the equation $D(x) = 0$.

\item $E_0(t)$ represents the spectral flow of the eigenvalue associated to the fundamental state of 
    the Hamiltonian $H_{\mr{A}}(t/T)$.

\end{enumerate}

According to the characteristics of the Hamiltonian $H_{\mr{A}}(t/T)$, only a finite adiabatic evolution $T$ time is necessary to obtain a state $\ket{n_0}$ which has a high probability to be the fundamental state of $H_{D}$ \cite{Kieu-2003}. The $\ket{n_0}$ state is observed from the emergence of a dominant probability in the set of probabilities $P_n(T)$. Once the  $\ket{n_0}$ state is obtained, its eigenvalue $E_0(T)$ is determined.  If  $E_0(T) =0$, the Diophantine equation $D(x) = 0$ has a solution in $\bbbz^{+2}$ and it is given by the value  $<n^2(T)>$. On the other hand, if $E_0(T) \neq 0$, then the equation $D(x) = 0$ doesn't have a solution in $\bbbz^{+2}$. The above description should be generalized for a Diophantine equation of $k$ variables.

\begin{figure}
\begin{center}
  \psset{unit=0.8cm}
    \begin{pspicture}(-6,0)(3,2)
      \pspolygon(-4,0)(-4,2)(0,2)(0,0)
      

      \rput(-2,1){${\displaystyle  \prod_{\tau =0}^{t-1} \frac{1-\frac{i}{2}H_{\mr{A}}(\tau/T)\delta t}{1+\frac{i}{2}H_{\mr{A}}(\tau/T)\delta t}}$}
      \rput(-6,1){\mbox{\large $\xrightarrow{H_D, \; H_{\mr{I}}, \; T, \; \delta t, \; \psi(0)}$}}
      \rput(2.5,1){\mbox{\large $\xrightarrow{\psi(t)}$} 
        $\begin{cases}
          P_n(t)
         \\
         <n^2(t)>
         \\
         E_0(t)
       \end{cases}$
}

    \end{pspicture}
\end{center}
\caption{Algorithm for the ISWHM}
\label{fig05}

\end{figure}

\section{The Simulation}
The idea is to simulate the algorithm described in Fig. \ref{fig05} in such a way that the realization of the adiabatic theorem for a big but finite $T$ time is observed. Since the algorithm operates upon a infinite dimensional Hilbert space, it is necessary to operate upon a truncated finite dimensional space to obtain the numeric solutions indicated by \eqref{cp-refesimu-90}. The truncation level $P$ means that we operate upon a space which canonical base is given by $\{\ket{1}, \dots, \ket{P} \}$.

For the carried out simulations, $\psi(0)$ is assumed as uniform and normalized, $H_{\mr{I}}$ is obtained from the matrix of ones, and $\delta t = 1$. Concretely, we carried out the simulation for the following simple Diophantine equations

\begin{align}
   x-16 &= 0 \enspace , \label{cp-refesimu-95} 
  \\
  x-7 &= 0 \enspace , \label{cp-refesimu-100}
  \\
 (x+1)(y+2) &= 12 \enspace , \label{cp-refesimu-105}
\end{align}
where \eqref{cp-refesimu-95} has a unique solution in $\bbbz^{+2}$, \eqref{cp-refesimu-100} doesn't have a solution in  $\bbbz^{+2}$ and \eqref{cp-refesimu-105} has a unique solution in $\bbbz^{+2} \times \bbbz^{+2}$.

\begin{figure}
   \begin{center}
     \subfigure[]{\label{fig10-a} \epsfig{figure=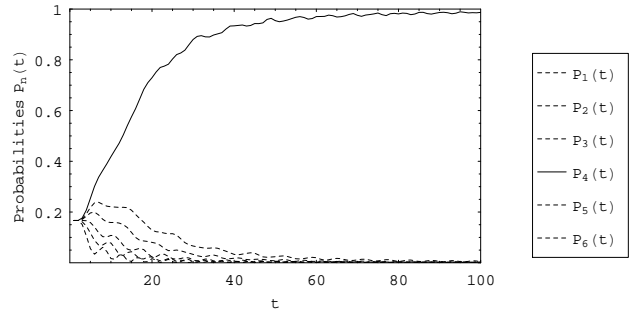,scale=0.5}}

     \subfigure[]{\label{fig10-b} \epsfig{figure=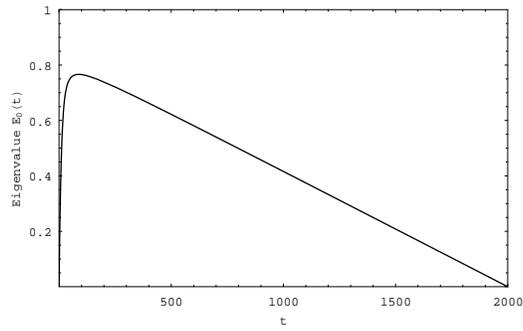,scale=0.4}}
     \subfigure[]{\label{fig10-c} \epsfig{figure=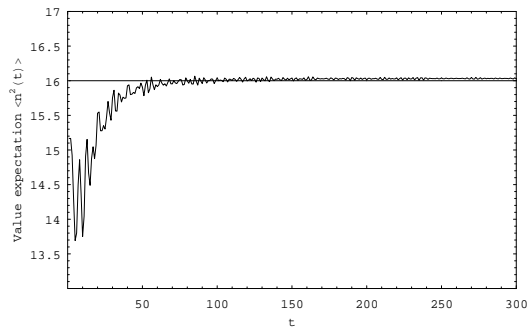,scale=0.4}}
     \caption{Simulation results for\eqref{cp-refesimu-95} with $T = 2000$ and $P = 6$}
     \label{fig10}
   \end{center}
 \end{figure}

For \eqref{cp-refesimu-95}, with $T  \approx 2000$ and $P=6$, it was determined that it had a solution. Figure \ref{fig10-a} points that rapidly $(t \approx 15)$ the state $\ket{n_0}= \ket{4}$ is the candidate to be the fundamental state $H_D$. Figure  \ref{fig10-b} indicates that this state effectively has the energy eigenvalue $E_0 =0$, and Fig. \ref{fig10-c} shows that the solution is given by $x = 16$. 

\begin{figure}
   \begin{center}
     \subfigure[]{\label{fig20-a} \epsfig{figure=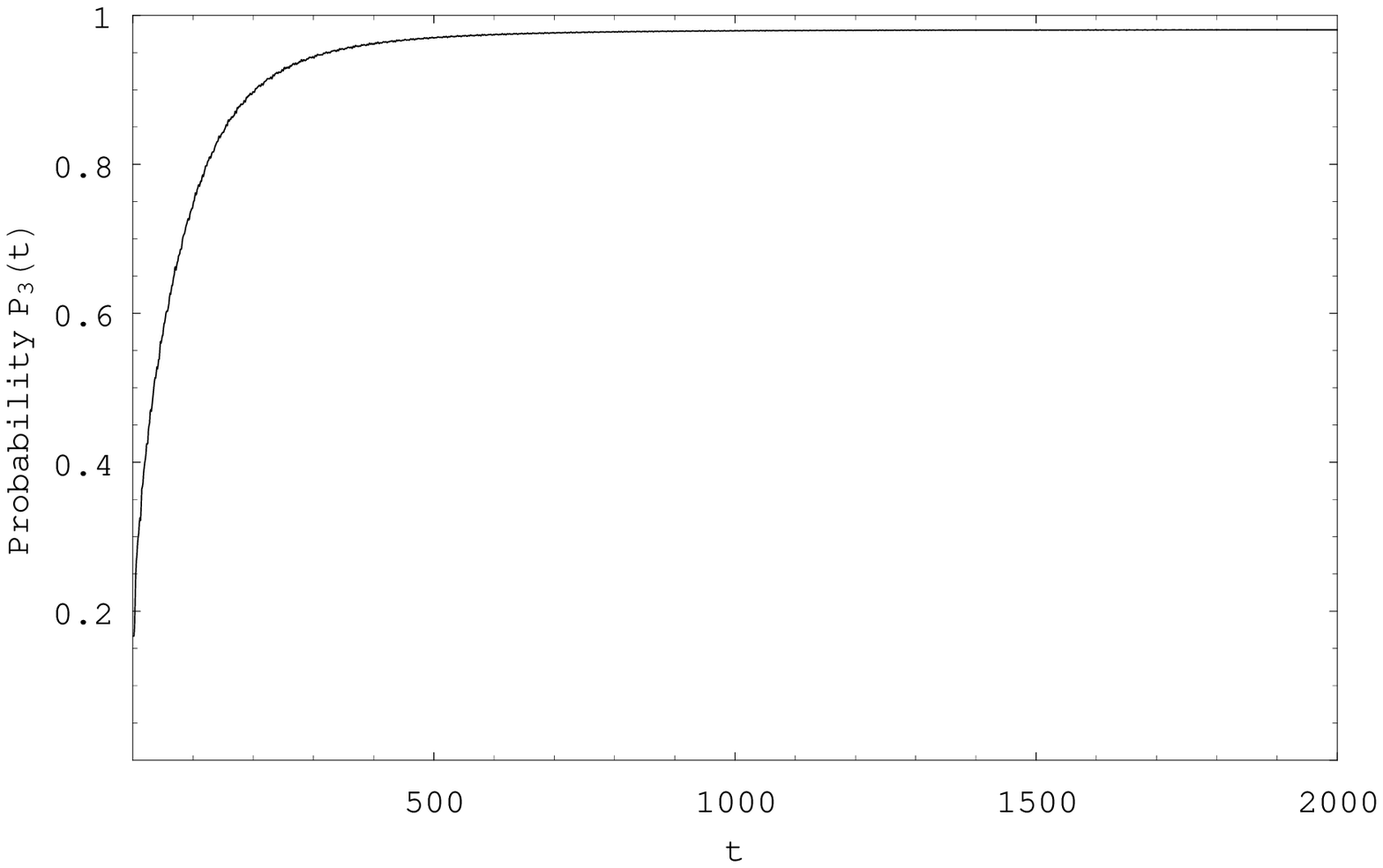,scale=0.4}}

     \subfigure[]{\label{fig20-b} \epsfig{figure=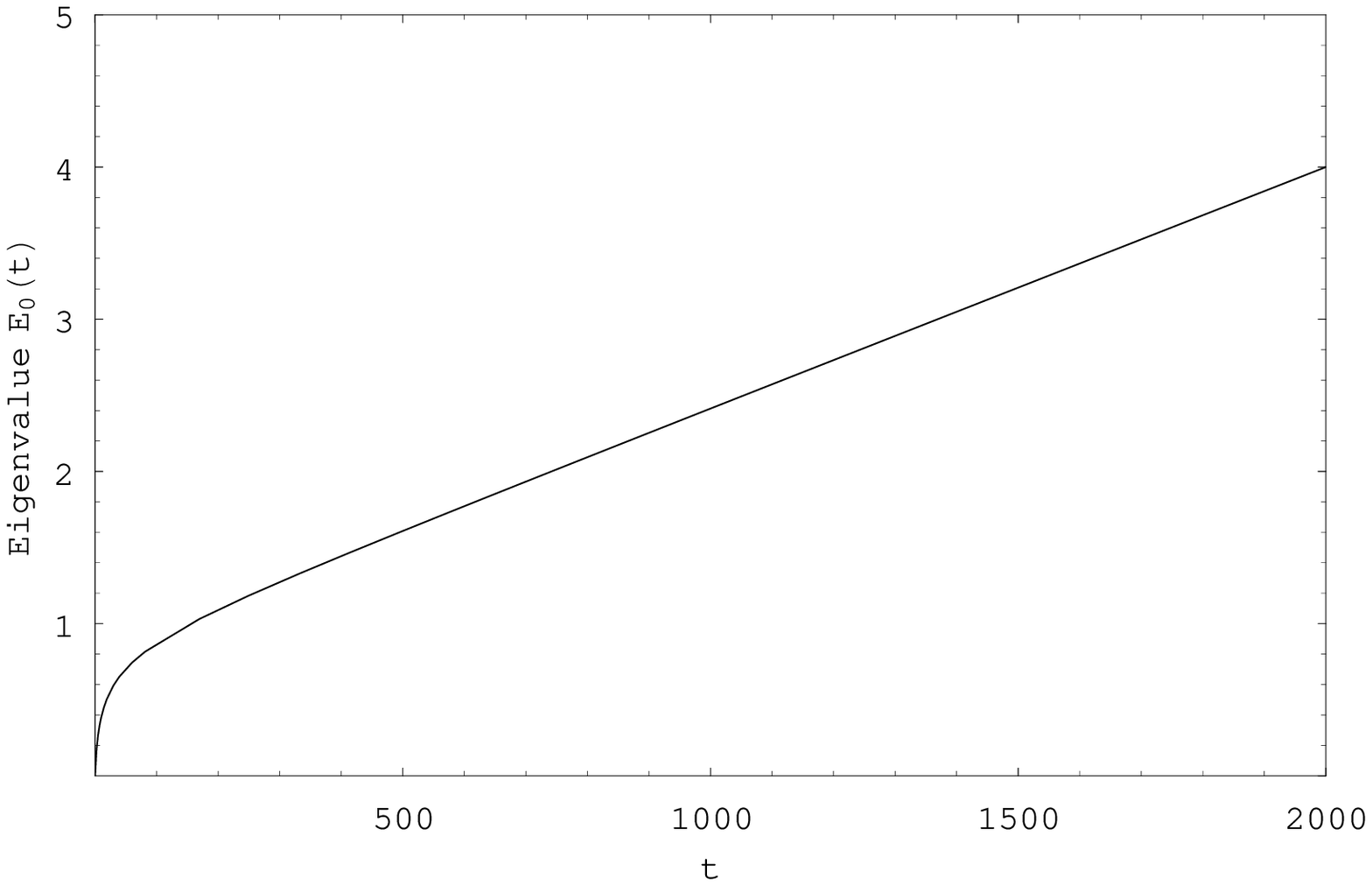,scale=0.4}}
     \caption{Simulation results for \eqref{cp-refesimu-100} with $T = 2000$ and $P = 6$}
     \label{fig20}
   \end{center}
 \end{figure}

For \eqref{cp-refesimu-100}, Fig. \ref{fig20-a} indicates that the probability that the state  $\ket{3}$ be the fundamental state of $H_D$ can be approximated to $1$ as much as we want. However, Fig. \ref{fig20-b}  points that  $E_0 \neq 0$, therefore, \eqref{cp-refesimu-100} doesn't have a solution in  $\bbbz^{+2}$.

\begin{figure}
   \begin{center}
     \subfigure[]{\label{fig30-a} \epsfig{figure=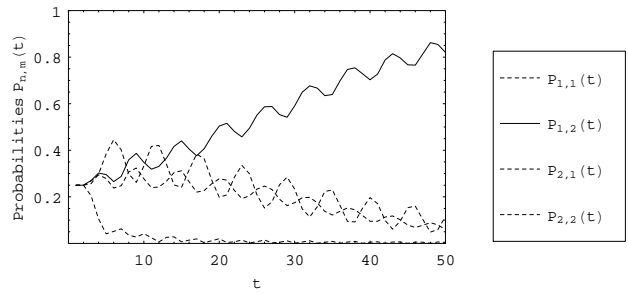,scale=0.5}}

     \subfigure[]{\label{fig30-b} \epsfig{figure=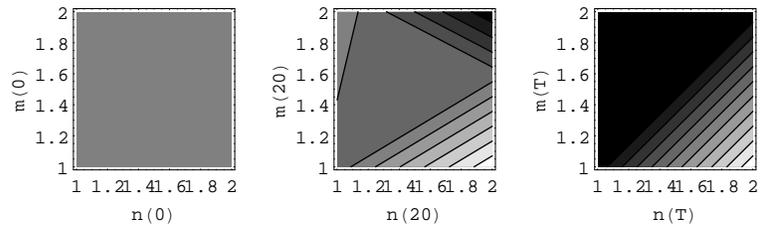,scale=0.6}}

     \subfigure[]{\label{fig30-c} \epsfig{figure=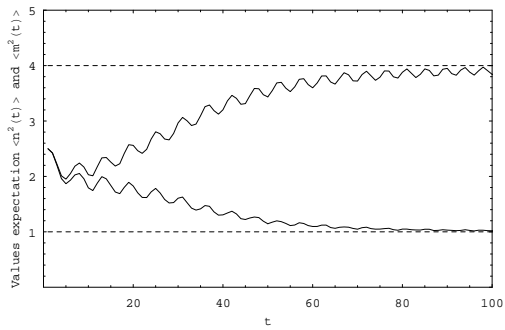,scale=0.4}}
     \caption{Simulation results for\eqref{cp-refesimu-105}  with $T = 1000$ and $P = 2$}
     \label{fig30}
   \end{center}
 \end{figure}

For \eqref{cp-refesimu-105}, Fig.  \ref{fig30-a} shows that the state $\ket{n_0,m_0}=\ket{1,2}$ tends to be the fundamental state, when $T = 1000$ and $P = 2$, and Fig.  \ref{fig30-b}  shows the contour curves for the codifying variables $n$ and $m$ at three different time moments. Figure  \ref{fig30-c} indicates that the equation solution is given by $x = 1$ and $y = 4$, supported by the fact that the spectral flow of eigenvalue $E_0(t)$ is similar to the one illustrated in Fig. \ref{fig10-b}.

\section{The Universality}
The universality of a quantum computation model, that is, its capability to carry out any operation that is carried out by a Turing machine, is established by its capability to generate a set of quantum gates, such that any unitary transformation $U(2^n)$, that is, any quantum gate that operates upon $n$-qubits, can be approximated with sufficient exactness by a quantum circuit that is only made of a finite number of gates of this set.

Henceforth, we will use the convention of  superindexes over the operators (quantum gates) and over the states (qubits). This  superindex will denote the dimension of Hilbert space upon which the operators act or upon which the states are defined. 

To demonstrate the universality of the ISWHM it is necessary and sufficient to construct three quantum gates: the controlled-NOT gate $C_{\mr{NOT}}^4$, the phase gate $\Phi_{\mr{\pi/4}}^2$  and Hadamard gate $H_{\mr{ad}}^2$ \cite{Boykin-Mor-Pulver-Roychowdhury-Vatan-1999,Velez-Sicard-Curty-2000}. For gates $C_{\mr{NOT}}^4$ and $\Phi_{\mr{\pi/4}}^2$, its matrix representation and the transformation carried out upon the base states of Hilbert space on which they act, are given by

\begin{multicols}{2}
\begin{equation*}
  C_{\mr{NOT}}^4 =\begin{pmatrix}
  1& 0 & 0 & 0 \\
  0 & 1 & 0 & 0 \\
  0 & 0 & 0 & 1 \\
  0 & 0 & 1 & 0
\end{pmatrix} \enspace,
\end{equation*}

\begin{equation}\label{cp-universalidad-01}
  \begin{split}
    \ket{00}^4  & \xrightarrow{C_{\mr{NOT}}^4}  \ket{00}^4\enspace ,
  \\
  \ket{01}^4 & \xrightarrow{C_{\mr{NOT}}^4}  \ket{01}^4 \enspace ,
  \\
  \ket{10}^4 & \xrightarrow{C_{\mr{NOT}}^4}  \ket{11}^4 \enspace ,
  \\
  \ket{11}^4 & \xrightarrow{C_{\mr{NOT}}^4}  \ket{10}^4 \enspace,
  \end{split}
\end{equation}
\end{multicols}

\begin{multicols}{2}
  \begin{equation*}
  \Phi_{\mr{\pi/4}}^2=\begin{pmatrix}
  1 & 0 \\
  0 & e^{i\frac{\pi}{4}}
\end{pmatrix} \enspace,
\end{equation*}

\begin{equation}\label{cp-universalidad-02}
  \begin{split}
    \ket{0}^2 & \xrightarrow{\Phi_{\mr{\pi/4}}^2} \ket{0}^2 \enspace,
    \\
    \ket{1}^2 & \xrightarrow{\Phi_{\mr{\pi/4}}^2} e^{i\pi/4}\ket{1}^2 \enspace .
    \end{split}
\end{equation}
\end{multicols}

In systems whose Hamiltonian operators are independent from the time, as it is the case of the infinite square well, the states of the system evolve according to the solution of Schrödinger equation of stationary states \cite{Galindo-Pascual-1978}

\begin{equation}\label{cp-universalidad-10}
  \ket{\psi(t)}=U(t)\ket{\psi(0)}=e^{-\frac{i}{\hbar}Ht}\ket{\psi(0)} \enspace,
\end{equation}
where $H$ is the Hamiltonian of the system defined in \eqref{cp-refesimu-15},  $U(t)$ is the evolution unitary operator and $\ket{\psi(0)}=\ket{n}^{\infty}$ is considered for the eigenstates of the infinite square well Hamiltonian. The $U(t)$ matrix elements are given by

\begin{equation}\label{cp-universalidad-15}
  U_{np}(t)=\exp \left (-i\frac{\hbar\pi^2 n^2t}{2mL^2} \right ) \delta_{np} \enspace ,
\end{equation}
where $\delta_{np}$  is Kronecker delta.

From different choices of the $t$ parameter in \eqref{cp-universalidad-15}, and different qubits coding, it is possible to find the evolutions corresponding to each of the mentioned quantum gates.

In order to build the phase gate $\Phi_{\mr{\pi/4}}^{\infty}$ in the ISWHM, based on the normalized eigenvectors of the phase gate $\Phi_{\mr{\pi/4}}^{2}$ \eqref{cp-universalidad-02}, we carry out a coding of the canonical base $\{\ket{0}^2,  \ket{1}^2 \}$ for a $1$-qubit in the canonical base $\{\ket{n}^{\infty}\}_{n  \in \bbbz^{+}}$, given by

\begin{align*}
    \ket{0}^2 &\xrightarrow{\mr{codification}}\ket{1}^{\infty}, 
    \\
    \ket{1}^2 &\xrightarrow{\mr{codification}}\ket{2}^{\infty} \enspace .
\end{align*}

According to \eqref{cp-universalidad-15} and for the time $t=\frac{2mL^2}{\hbar\pi^{2}}\phi$, where $\phi \in [-\pi,\pi)$, we obtain a quantum gate $\Gamma^{\infty}(\phi)$ whose components are given by

\begin{equation}\label{cp-universalidad-40}
  \Gamma^{\infty}_{np}(\phi) = \left [ \exp \left ( -i\phi \right ) \exp \left (-i\phi(n^2-1) \right ) \right ]\delta_{np} \enspace ,
\end{equation}
where term $\exp \left ( -i\phi \right )$ hasn't physical meaning.

With \eqref{cp-universalidad-40}  it is possible to control, at least theoretically, the ISWHM, until the implementation of the transformation that accounts for the phase gate $\Phi_{\mr{\pi/4}}^{\infty}$, given by

\begin{multicols}{2}
\begin{equation*}
  \begin{split}
    \Phi_{\mr{\pi/4}}^{\infty} &= \Gamma^{\infty}(-\pi/12)
    \\
    &= U \left (\frac{2mL^2}{\hbar\pi^{2}} \left(-\frac{\pi}{12} \right) \right) \enspace ,
\end{split}
\end{equation*}

\begin{equation*}
  \begin{split}
    \ket{1}^{\infty} & \xrightarrow{\Phi_{\mr{\pi/4}}^{\infty}}\ket{1}^{\infty} \enspace , 
\\
\ket{2}^{\infty}&\xrightarrow{\Phi_{\mr{\pi/4}}^{\infty}}e^{i\frac{\pi}{4}}\ket{2}^{\infty} \enspace .
\end{split}
\end{equation*}
\end{multicols}

On the other hand, in order to build the $C_{\mr{NOT}}^{\infty}$ gate in the ISWHM, based on the normalized eigenvectors of the  $C_{\mr{NOT}}^{2}$ gate \eqref{cp-universalidad-01}, we code the base 

\begin{equation*}
\{\ket{00}^4,\ket{01}^4, \frac{\ket{10}^4+\ket{11}^4 }{\sqrt{2}},\frac{\ket{11}^4-\ket{10}^4 }{\sqrt{2}}\}
\end{equation*}
for a $2$-qubit in the canonical base $\{\ket{n}^{\infty}\}_{n  \in \bbbz^{+}}$ by means of 

\begin{align*}
\ket{00}^4  &\xrightarrow{\mr{codification}} \ket{2}^{\infty}  \enspace, 
\\
 \ket{01}^4  & \xrightarrow{\mr{codification}} \ket{4}^{\infty} \enspace , 
\\
\frac{\ket{10}^4+\ket{11}^4 }{\sqrt{2}} & \xrightarrow{\mr{codification}} \frac{\ket{6}^{\infty}+\ket{1}^{\infty} }{\sqrt{2}} \enspace , 
\\
\frac{\ket{11}^4-\ket{10}^4 }{\sqrt{2}}  & \xrightarrow{\mr{codification}} \frac{\ket{6}^{\infty}-\ket{1}^{\infty}
 }{\sqrt{2}} \enspace .
\end{align*}

According to \eqref{cp-universalidad-15}, in  $t=\frac{2mL^2}{\hbar\pi}$ we obtain the $C_{\mr{NOT}}^{\infty}$ gate given by 

\begin{multicols}{2}
\begin{equation*}
  \begin{split}
    C_{\mr{NOT}}^{\infty} &= U \left ( \frac{2mL^2}{\hbar\pi^{2}} \right ) \enspace,
    \\
\\
\\
\end{split}
\end{equation*}

\begin{equation*}
  \begin{split}
    \ket{2}^{\infty}  &\xrightarrow{C_{\mr{NOT}}^{\infty}} \ket{2}^{\infty} \enspace, 
\\
 \ket{4}^{\infty} & \xrightarrow{C_{\mr{NOT}}^{\infty}} \ket{4}^{\infty} \enspace, 
\\ 
\frac{\ket{6}^{\infty}+\ket{1}^{\infty}
 }{\sqrt{2}}& \xrightarrow{C_{\mr{NOT}}^{\infty}} \frac{\ket{6}^{\infty}-\ket{1}^{\infty} }{\sqrt{2}} \enspace, 
\\
\frac{\ket{6}^{\infty}-\ket{1}^{\infty}}{\sqrt{2}}&\xrightarrow{C_{\mr{NOT}}^{\infty}}\frac{\ket{6}^{\infty}+\ket{1}^{\infty}}{\sqrt{2}} \enspace.
\end{split}
\end{equation*}

\end{multicols}

In a similar fashion, the $H_{\mr{ad}}^{\infty}$  gate is built in the ISWHM by making a coding of the $\{\ket{0}^2,  \ket{1}^2 \}$ base in the $\{\ket{n}^{\infty}\}_{n  \in \bbbz^{+}}$  base, from the normalized eigenvectors of the $H_{\mr{ad}}^{2}$ Hadamard gate, and by selecting an adequate value of the $t$ parameter in \eqref{cp-universalidad-15}.

Although from a theoretical point of view we got the universality of the ISWHM, this model doesn't turn out to be adequate from its physical implementation point of view due to the realization of the mentioned quantum gates in different times and due to the change in the coding of the base of the qubits for each gate.

\section{Conclusions}
It is quite surprising that a quantum computation model over such a \emph{simple} physical referent like the infinite square well have hypercomputation characteristics as it was demonstrated for the ISWHM. In fact, and according to the obtained universality results, the ISWHM constitutes itself (at least from a theoretical perspective) as a super-Turing computational model. 

On the other hand, the success obtained by choosing a physical referent different from the one selected by Kieu opens the possibility of obtaining new hypercomputation models supported on the quantum computation. Due to the nature of their theoretical construction, these models may be the best candidates to a possible physical realization or implementation of a hypercomputer. Be this possible, such hypercomputer would broaden the conception of the term `computable', not exclusively from the theoretical perspective but also from the practical perspective.

\section*{Aknowledgments}
We would like to thank Tien D. Kieu for discussions and feedback. This research was supported by EAFIT University under Research Project No. $PY0204$.

\end{document}